\documentclass{article}

\usepackage{eusipco2010}                       
\usepackage{graphicx}

\title{A Comparative Evaluation of Pitch Modification Techniques}

\name {Thomas Drugman, Thierry Dutoit}

\address{Facult\'e Polytechnique de Mons - TCTS Lab\\
University of Mons, 7000, Mons, Belgium\\
phone: + (32) 65 37 47 49, fax: + (32) 65 37 47 29, email: thomas.drugman@umons.ac.be \\
web: http://tcts.fpms.ac.be/~drugman/}

\begin{document}

\maketitle

\begin{abstract}
This paper addresses the problem of pitch modification, as an important module for an efficient voice transformation system. The Deterministic plus Stochastic Model of the residual signal we proposed in a previous work is compared to TDPSOLA, HNM and STRAIGHT. The four methods are compared through an important subjective test. The influence of the speaker gender and of the pitch modification ratio is analyzed. Despite its higher compression level, the DSM technique is shown to give similar or better results than other methods, especially for male speakers and important ratios of modification. The DSM turns out to be only outperformed by STRAIGHT for female voices.
\end{abstract}

\section{Introduction}\label{sec:intro}

Voice transformation refers to the various modifications one may apply to the sound produced by a person such that it is perceived as uttered by another speaker \cite{Stylianou-Survey}. These modifications encompass various properties of the speech signal such as prosodic, vocal tract-based as well as glottal characteristics. Although all these features should be taken into account in an efficient voice transformation system, this study only focuses on pitch modifications, as pitch is an essential aspect in the way speech is perceived. More precisely, the main goal of this paper is to compare the Deterministic plus Stochastic Model (DSM) of the residual signal we proposed in \cite{DSM} to the main state-of-the-art techniques of pitch modification.

The paper is structured as follows. Section \ref{sec:Methods} gives a brief overview on the methods considered in this study, namely: the DSM of the residual signal \cite{DSM}, the Time-Domain Pitch-Synchronous Overlap-Add technique (TDPSOLA, \cite{TDPSOLA}), the Harmonic plus Noise Model of speech (HNM, \cite{HNM}) and STRAIGHT \cite{STRAIGHT}. In Section \ref{sec:Exp} these methods are compared through a subjective evaluation regarding their pitch modification capabilities. Finally Section \ref{sec:conclu} concludes and discusses in depth the main observations drawn from the results.

\section{Methods for Pitch Modification}\label{sec:Methods}

Various approaches for pitch modification have already been proposed in the literature. Some of them are based on a parametric modeling (HNM \cite{HNM}, STRAIGHT \cite{STRAIGHT}, ARX-LF \cite{ARX-LF}), or on a phase vocoder \cite{PhaseVocoder}, \cite{Depalle}, while others rely on a non-parametric representation (TDPSOLA \cite{TDPSOLA}). This section briefly presents the methods that will be compared in Section \ref{sec:Exp}. In Section \ref{ssec:DSM}, the Deterministic plus Stochastic Model (DSM) of the residual signal we proposed in \cite{DSM} is described. The three next subsections (Sections \ref{ssec:TDPSOLA}, \ref{ssec:HNM} and  \ref{ssec:STRAIGHT}) respectively review the TDPSOLA, HNM and STRAIGHT algorithms. For information, the footprint of each method is presented in number of parameters/second, giving an idea of their compression level. 

\subsection{Deterministic plus Stochastic Model of the Residual Signal}\label{ssec:DSM}
\begin{figure}[!ht]
  \centering
  \includegraphics[width=0.45\textwidth]{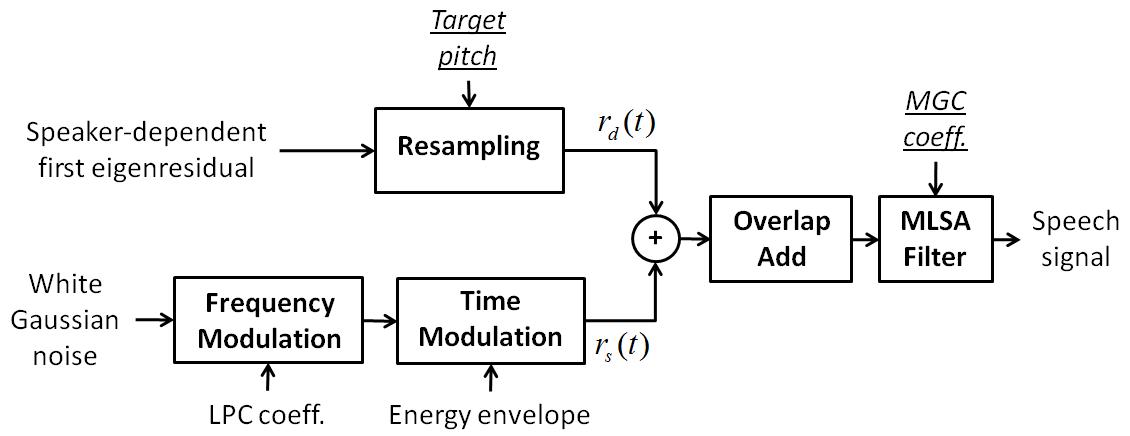}
  \caption{Workflow of the DSM vocoder. Inputs are the target pitch values and the MGC coefficients.}
  \label{fig:DSMvocoder}    
\end{figure}

In \cite{DSM}, we proposed a Deterministic plus Stochastic Model (DSM) of the residual signal. This approach was reported to significantly improve the quality delivered by the basic HMM-based speech synthesizer. The workflow of the DSM vocoder is presented in Figure \ref{fig:DSMvocoder}. The DSM consists of the superposition of a deterministic $r_d(t)$ and stochastic $r_s(t)$ components of the residual. These components act in two distinct spectral bands delimited by the maximum voiced frequency $F_m$ (as introduced in the HNM \cite{HNM}). For unvoiced frames, $F_m=0$ and a simple white noise is used as excitation. For voiced frames, $F_m$ is fixed to $4 kHz$ (although one could think of using a static value depending on the considered voice, or even of a dynamic approach as in the HNM \cite{HNM}).

\begin{figure}[!ht]
  \centering
  \includegraphics[width=0.45\textwidth]{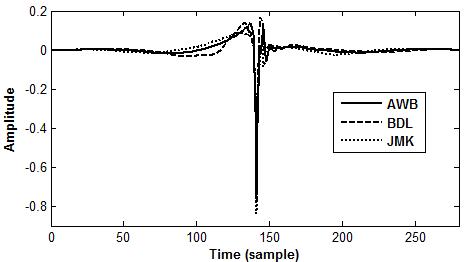}
  \caption{First eigenresidual for three male speakers of the CMU ARCTIC database.}
  \label{fig:Eigenresiduals}  
\end{figure}

The deterministic part $r_d(t)$ relies on a specific speaker-dependent waveform called \emph{eigenresidual}. This eigenresidual results from the following procedure. A speaker-dependent speech corpus is analyzed and Mel-Generalized Cepstral (MGC, \cite{MGC}) coefficients are extracted. Residual signals are then obtained by inverse filtering. Glottal Closure Instants (GCIs) are accurately located on this signal using the method described in \cite{Drugman-GCI}. Residual frames are then isolated by applying a GCI-centered two pitch period-long windowing. Since the resulting frames have different lengths, a resampling step (by interpolation/decimation) on a fixed number of points is required. Note that this normalization length should respect some criterion avoiding high-frequency information loss, as explained in \cite{DSM}. Frames are finally normalized in energy. All this process of GCI-synchronization and prosody normalization is required in order to ensure that the resulting residual frames are suited for a common modeling. In this way, a coherent dataset containing thousands of pitch-synchronous normalized residual frames is extracted. The eigenresidual is then defined as the first eigenvector obtained by computing PCA on this large dataset. Note that at synthesis time, as shown in Figure \ref{fig:DSMvocoder}, a step of resampling to the target pitch value  is required since the eigenresidual was pitch-normalized. Figure \ref{fig:Eigenresiduals} illustrates the typical waveform of the resulting eigenresidual for three male speakers of the CMU ARCTIC database. Interestingly, a strong similarity with models of the glottal flow (such as the LF model \cite{LF}), mainly during the glottal open phase, can be noticed.

As shown in Figure \ref{fig:DSMvocoder}, the stochastic part $r_s(t)$ consists of a white Gaussian noise filtered by an auto-regressive model beyond $F_m$, and whose time structure is controled by an envelope (modeling the natural pitch-synchronous noise modulation \cite{Hermes}). In this work, we employed the basic triangular window used in the HNM \cite{HNM}, although it was shown that other parametric or non-parametric envelopes could lead to a higher perceptual quality \cite{Stylianou-Envelope}. It is worth noting that this energy envelope, which modulates the time evolution of the stochastic part, is GCI-synchronous and its length is adapted to the pitch period. As for the filter used for the frequency modulation, it is estimated as the Linear Predictive modeling of the averaged high-frquency content (beyond $F_m$) computed on the considered residual dataset.

The resulting residual frames are finally overlap-added and filtered by the MGC coefficients to get the synthetic speech signal. We used as inputs of the workflow 25 MGC parameters for the vocal tract, and only $F_0$ for the excitation, all other data being pre-computed on the speaker-dependent database. These features are extracted every 5 ms which leads to a 5200 parameters/s vocoder.

\subsection{Time-Domain Pitch-Synchronous Overlap-Add}\label{ssec:TDPSOLA}
The TDPSOLA technique \cite{TDPSOLA} is probably the most famous non-parametric approach for pitch modification. According to this method, pitch-synchronous speech frames whose length is a multiple of the pitch period are duplicated or eliminated. It is in this way assumed that the pitch can be modified while keeping the vocal tract characteristics unchanged. In our implementation we considered two pitch period-long speech frames centered on the GCIs. GCI positions were located by the method described in \cite{Drugman-GCI}, providing a high-quality phase synchronization. As this technique is based on the speech waveform itself (sampled at 16 kHz in our experiments), 16000 values/s are necessary.

\subsection{Harmonic plus Noise Model}\label{ssec:HNM}

The Harmonic plus Noise Model (HNM, \cite{HNM}) assumes the speech signal to be composed of a harmonic part and a noise part. The harmonic part accounts for the quasi-periodic component of the speech signal while the noise part accounts for its non-periodic components (e.g., fricative or aspiration noise, etc.). The two components are separated in the frequency domain by a time-varying parameter, referred to as maximum voiced frequency $F_m$. The lower band of the spectrum (below $F_m$) is assumed to be represented solely by harmonics while the upper band (above $F_m$) is represented by a modulated noise component. In this study, we used the HNM algorithm with its default options. Since the number of harmonics (and consequently of parameters) is different regarding $F_0$ and $F_m$, the bitrate may vary across speakers and sentences. In average, we found that around 10000 parameters were necessary for coding 1s of speech.



\subsection{STRAIGHT}\label{ssec:STRAIGHT}
STRAIGHT is a well-known vocoding system \cite{STRAIGHT} which showed its ability to produce high-quality voice manipulation and was successfully incorporated into HMM-based speech synthesis. STRAIGHT is basically based on both a source information extractor as well as a smoothed time-frequency representation \cite{STRAIGHT}. In this work, we employed the version publicly available in \cite{STRAIGHT-Web} with its default options. In this implementation, the algorithm extracts every 1 ms: the pitch, aperiodic components of the excitation (513 coeff.) and a representation of the smoothed spectrogram (513 coeff.). This leads to a high-quality vocoder using a bit more than 1 million parameters/s.


\section{Experiments}\label{sec:Exp}

In this part, methods presented in Section \ref{sec:Methods} are evaluated on 3 male (AWB, BDL and JMK) and 2 female (CLB and SLT) speakers from the CMU ARCTIC database \cite{ARCTIC}. For each speaker, the three first sentences of the database were synthesized using the four techniques, and this for 5 pitch modification ratios: 0.5, 0.7, 1, 1.4 and 2. This leads to a total set containing 300 sentences. The DSM technique was compared to the three other approaches (TDPSOLA, HNM and STRAIGHT) through a Comparative Mean Opinion Score (CMOS) test composed of 30 pairwise sentences chosen randomly among the total set. 27 people (mainly naive listeners) participated to the test. For each sentence they were asked to listen to both versions (randomly shuffled) and to attribute a score according to their overall preference. The CMOS values range on a gradual scale varying from -3 (meaning that DSM is much worse than the other technique) to +3 (meaning the opposite). A score of 0 is given if both versions are found to be equivalent. It is worth noting that, due to the unavailability of a ground truth reference of how a sentence whose pitch has been modified by a given factor should sound, participants were asked to score according to their overall appreciation of the different versions. These scores then reflect both the quality of pitch modification, as well as the possible artifacts that the different signal representations may generate.

\begin{figure}[!ht]
  \centering
  \includegraphics[width=0.45\textwidth]{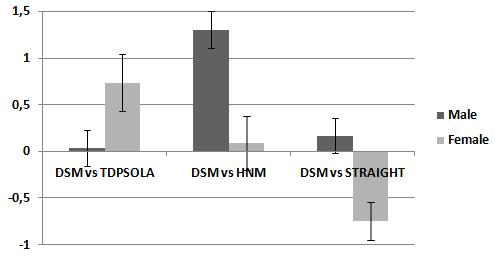}
  \caption{CMOS results together with their 95\% confidence intervals for the three comparisons and for both male and female speakers.}
  \label{fig:CMOS_Gender}    
\end{figure}

Figure \ref{fig:CMOS_Gender} displays the CMOS results with their 95\% confidence intervals for the three comparisons and according to the gender of the speaker. For male voices, it can be noticed that DSM gives scores similar to TDPSOLA, while its advantage over HNM, and STRAIGHT in a lesser extent, is appreciable. For female speakers, the tendency is inversed. DSM is comparable to HNM while it is superior to TDPSOLA. Albeit for such comparative subjective tests transitional properties can not be assumed to hold, it however seems that STRAIGHT outperforms all other techniques for female voices. It is also worth noticing that the degradation of DSM with regard to STRAIGHT for female speakers is done at the expense of a high gain of compression and complexity. Depending on the considered application, the choice of one of the compared method should then result from a trade-off between these latter criteria (i.e speech quality vs compression rate).

\begin{figure*}[!ht]
  \centering
  \includegraphics[width=0.95\textwidth]{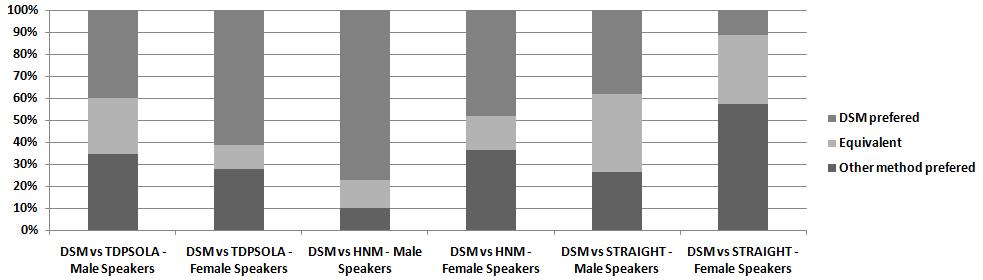}
  \caption{Preference scores for the three comparisons and for both male and female speakers.}
  \label{fig:PrefTest}  
\end{figure*}

In Figure \ref{fig:PrefTest} the preference scores for both male and female speakers can be found. Although somehow redundant with the previous results, this figure conveys information about the percentage of preference for a given method and about the ratio of indifferent opinions. Interestingly it can be noted that DSM was in general prefered to other methods, except for female speakers where STRAIGHT showed a clear advantage. In \cite{ARX-LF}, authors compared an improved ARX-LF framework to TDPSOLA and HNM through a small preference test. Even though these results are obviously not extrapolable, the preference scores they obtain are stronlgy similar to ours, except for the comparison with TDPSOLA on female voices where ARX-LF was shown to be inferior.

\begin{figure*}[!ht]
  \centering
  \includegraphics[width=0.95\textwidth]{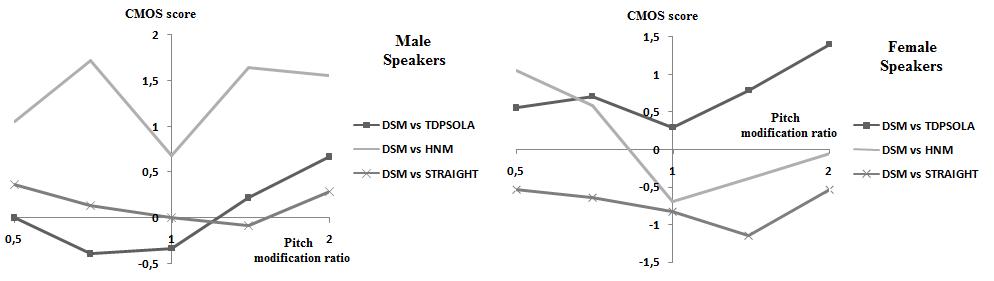}
  \caption{Evolution of the CMOS results with the pitch modification ratio for both male and female speakers.}
  \label{fig:PitchRate}  
\end{figure*}

Finally, the evolution of the performance with the pitch modification ratio is analyzed in Figure \ref{fig:PitchRate}. As a reminder, the higher the CMOS score, the more DSM was prefered regarding the method to which it was compared. A positive (negative) value means that, in average, the DSM (the other method) was prefered. Interestingly, it can be observed that in general plots tend to exhibit a minimum in 1 (where no pitch modification was applied) and go up around this point. This implies that the relative performance of DSM over other techniques increases as the pitch modification ratio is important. This was expected regarding the comparison with TDSPOLA, but the same observation seems to hold for STRAIGHT, and for HNM (though in a lesser extent for male voices). Note that in our implementation of TDPSOLA, a GCI-synchronous overlap-add was performed even when no pitch modification was required. This may explain why, for female voices (for which GCIs are known to be difficult to be precisely located), listeners slighlty prefered DSM over TDPSOLA, even without pitch modification.

\section{Conclusions and discussion}\label{sec:conclu}
This paper proposed a comparison between the DSM of the residual signal and 3 other well-known methods of pitch modification: TDPSOLA, HNM and STRAIGHT. An important subjective test allowed a comparative evaluation of these 4 techniques. From this study, several conclusions can be drawn:

\begin{itemize}

\item Interestingly the DSM approach, despite its small footprint, gives similar or better results than other state-of-the-art techniques. Its efficiency probably relies on its ability to implicitly capture and process the essential of the phase information via the eigenresidual. The pitch-dependent resampling operations involved in its process indeed preserve the most important glottal properties (such as the open quotient and asymmetry coefficient). Nevertheless a degradation for female speakers is noticed. This can be mainly explained by the fact that the spectral envelope we used may contain pitch information. Although this effect can be alleviated by the use of Mel-Generalized Cepstral (MGC, \cite{MGC}) coefficients instead of the traditional LPC modeling (since MGCs make use of a warped frequency axis), it may still occur for high-pitched voices where the risk of confusion between $F0$ and the first formant $F1$ is more important. After pitch modification, this effect leads to detrimental source-filter interactions, giving birth to some audible artefacts. Note that this effect is almost completely avoided with STRAIGHT, as this method makes use of a time-frequency smooth representation of the spectral envelope \cite{STRAIGHT}. Reducing this drawback within the DSM framework is the object of ongoing work, possibly by applying a GCI-synchronous spectral analysis instead of achieving it in an asynchronous way.

\item The results we obtained for HNM corroborate the conclusions from \cite{ARX-LF} and \cite{Kim}. In \cite{Kim}, the observation that sinusoidal coders produce higher quality speech for female speakers than for male voices is justified by the concept of critical phase frequency, below which phase information is perceptually irrelevant. Note also that we used the HNM algorithm with its default options. In this version, we observed that the quality of the HNM output was strongly affected for some voices by a too low estimation of the maximum voiced frequency. This led to an unpleasant predominance of noise in the speech signal. Fixing the maximum voiced frequency to a constant value (as in the DSM technique) could lead to a relative improvement for these problematic voices.

\item The degradation of TDPSOLA for female speakers is probably due to the difficulty in obtaining accurate pitch marks for such voices. This results in inter-frame phase incoherences, degrading the final quality. Besides note that TDPSOLA requires the original speech waveform as input and is then not suited for parametric speech synthesis.

\item It turns out from this study and the one exposed in \cite{ARX-LF} that approaches based on a source-filter representation of speech lead to the best results. This is possible since these techniques process the vocal tract and the glottal contributions independently. Among these methods, STRAIGHT gives in average the best results but requires heavy computation load. The DSM and the improved ARX-LF technique proposed in \cite{ARX-LF} seem to lead to a similar quality. Note that STRAIGHT and the DSM were successfully integrated into a HMM-based speech synthesizer (respectively in \cite{STRAIGHT2} and \cite{DSM}). In \cite{Cabral}, it was also proposed to incorporate the traditional ARX-LF model in a statistical parametric synthesizer. Although an improvement regarding the basic baseline was reported, it seems that this latter is less significant than it was achieved by STRAIGHT and DSM. It is then clear that the good quality obtained in \cite{ARX-LF} and \cite{ARX-LF2} with the improved ARX-LF method is reached thanks to the modeling of the LF-residual (i.e the signal obtained after removing the LF contribution in the excitation). This is possible in an analysis-synthesis task (where the target LF-residual is available), but was not yet carried out in speech synthesis. Finally note that the ARX-LF approach has the flexibility to potentially produce easy modifications of voice quality or emotion (\cite{ARX-LF2}, \cite{Cabral}) since it relies directly on a paramectric model of the glottal flow (which is not the case for the DSM and STRAIGHT techniques).

\end{itemize}

\section{Acknowledgments}\label{sec:Acknowledgments}

Thomas Drugman is supported by the ``Fonds National de la Recherche Scientifique'' (FNRS). The authors would like to thank Prof. Stylianou for providing the HNM code, as well as Prof. Kawahara for making the version of STRAIGHT publicly available. Authors are also deeply indebted to Alexis Moinet and Geoffrey Wilfart for their precious help.

\vspace{20pt}



\end{document}